\documentclass[sigconf]{acmart}
\usepackage{bm}
\usepackage{hyperref}

\AtBeginDocument{%
  \providecommand\BibTeX{{%
    \normalfont B\kern-0.5em{\scshape i\kern-0.25em b}\kern-0.8em\TeX}}}

\copyrightyear{2021}
\acmYear{2021}
\setcopyright{acmcopyright}
\acmConference[CHI '21]{CHI Conference on Human Factors in Computing Systems}{May 8--13, 2021}{Yokohama, Japan}
\acmBooktitle{CHI Conference on Human Factors in Computing Systems (CHI '21), May 8--13, 2021, Yokohama, Japan}
\acmPrice{15.00}
\acmDOI{10.1145/3411764.3445115}
\acmISBN{978-1-4503-8096-6/21/05}

\begin{document}

\title{Optimal Action-based or User Predict\-ion-based Haptic Guidance: Can You Do Even Better?}

\author{Hee-Seung Moon}
\affiliation{%
  \institution{Yonsei University}
  \city{Incheon}
  \country{Korea}
  \postcode{21983}}
\email{hs.moon@yonsei.ac.kr}

\author{Jiwon Seo}
\affiliation{%
  \institution{Yonsei University}
  \city{Incheon}
  \country{Korea}
  \postcode{21983}}
\email{jiwon.seo@yonsei.ac.kr}

\begin{abstract}
  The recently advanced robotics technology enables robots to assist users in their daily lives. Haptic guidance (HG) improves users' task performance through physical interaction between robots and users. It can be classified into optimal action-based HG (OAHG), which assists users with an optimal action, and user prediction-based HG (UPHG), which assists users with their next predicted action. This study aims to understand the difference between OAHG and UPHG and propose a combined HG (CombHG) that achieves optimal performance by complementing each HG type, which has important implications for HG design. We propose implementation methods for each HG type using deep learning-based approaches. A user study (n=20) in a haptic task environment indicated that UPHG induces better subjective evaluations, such as naturalness and comfort, than OAHG. In addition, the CombHG that we proposed further decreases the disagreement between the user intention and HG, without reducing the objective and subjective scores.
\end{abstract}

\begin{CCSXML}
<ccs2012>
   <concept>
       <concept_id>10003120.10003121</concept_id>
       <concept_desc>Human-centered computing~Human computer interaction (HCI)</concept_desc>
       <concept_significance>500</concept_significance>
       </concept>
   <concept>
       <concept_id>10003120.10003121.10003125.10011752</concept_id>
       <concept_desc>Human-centered computing~Haptic devices</concept_desc>
       <concept_significance>300</concept_significance>
       </concept>
   <concept>
       <concept_id>10003120.10003121.10003122.10003332</concept_id>
       <concept_desc>Human-centered computing~User models</concept_desc>
       <concept_significance>300</concept_significance>
       </concept>
   <concept>
       <concept_id>10010147.10010257</concept_id>
       <concept_desc>Computing methodologies~Machine learning</concept_desc>
       <concept_significance>300</concept_significance>
       </concept>
 </ccs2012>
\end{CCSXML}

\ccsdesc[500]{Human-centered computing~Human computer interaction (HCI)}
\ccsdesc[300]{Human-centered computing~Haptic devices}
\ccsdesc[300]{Human-centered computing~User models}
\ccsdesc[300]{Computing methodologies~Machine learning}

\keywords{Haptic guidance, Physical human--robot interaction (pHRI), User adaptation, Deep learning}

\begin{teaserfigure}
  \includegraphics[width=\textwidth]{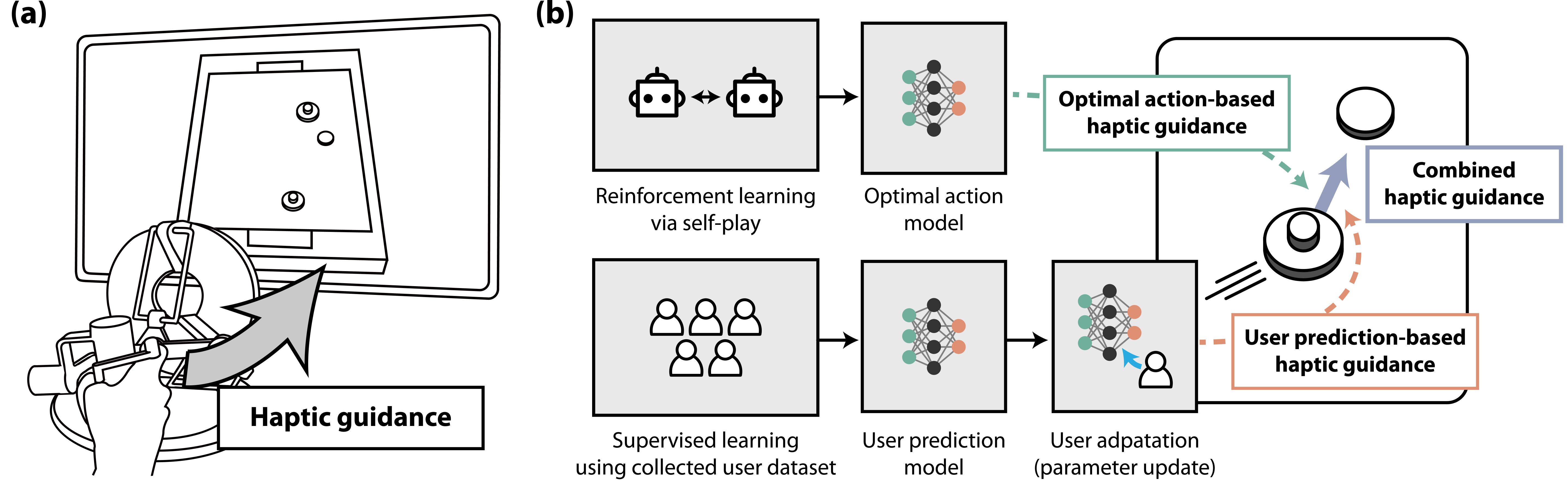}
  \caption{Overview of this research. (a) We built a virtual air hockey task environment controlled by a haptic device. Haptic guidance that assists a user's task performance is delivered through this device. (b) We implemented optimal action-based and user prediction-based haptic guidance using deep learning-based approaches, proposed a combined haptic guidance for better performance, and experimentally compared them.}
  \label{fig:01}
\end{teaserfigure}

\maketitle

\section{Introduction}
With the recent advancements in artificial intelligence and robotics technology, it is becoming increasingly common for users to be assisted by robots or computing machineries. In the context of physical human–-robot interaction (pHRI), haptic modality has a high use potential; haptic feedback can directly pass through the neuromuscular system without going through a high-level recognition process~\cite{abbink2010neuromuscular}; therefore, a user can respond faster than when other modalities (e.g., visual feedback) are used. Recently, the haptic guidance (HG) system, also called the haptic shared control system, has been accepted as a promising approach for human–-machine interfaces or pHRI situations~\cite{abbink2012haptic}. The HG system is defined as a method in which the control input determined through physical interaction between the force exerted by a human operator and the guiding force of a robot is applied to the target system~\cite{abbink2010neuromuscular}. Compared to the previous support systems using haptic cueing~\cite{lindeman2005effectiveness, di2019haptic} or input-mixing shared control~\cite{dragan2013policy}, an advantage of the HG system is that the user can not only recognize the intention of the robot (e.g., direction and strength), but also choose to what extent this intention to be reflected. For example, a user usually follows the robot's guidance, but when his/her choice is necessary, he/she can apply a force that exceeds the HG to perform the desired control.

HG technologies presented in earlier studies can be classified into two categories according to their design method: \textit{optimal action-based haptic guidance} (OAHG) and \textit{user prediction-based haptic guidance} (UPHG). OAHG is designed to convey the optimal movement for performing a task in the current state (e.g., guiding the user toward the center of the road in a steering task~\cite{forsyth2005predictive, mulder2008effect, wang2017effect} or through the movement of a skilled expert in a peg-in-hole task~\cite{perez2016using}). Meanwhile, UPHG is designed to provide proactive guidance in the direction the user intends to move based on their behavior prediction. For example, in a steering task, it guides the users to their individually preferred courses rather than the center of the road~\cite{de2015effect}. Both types of HG systems have been proven to have several positive effects in terms of task performance improvement~\cite{forsyth2005predictive, mulder2008effect, de2015effect, bluteau2008haptic, kuiper2016evaluation, hernandez2015synthesizing, zahedi2017gesture}, user workload reduction~\cite{mulder2008effect, kuiper2016evaluation, hernandez2015synthesizing}, and user subjective satisfaction~\cite{forsyth2005predictive, kuiper2016evaluation, hernandez2015synthesizing} in various recent studies.

While most of the previous studies deal with the design methods and effects of OAHG and UPHG, very few studies have clearly compared OAHG and UPHG. The two HGs are expected to have different characteristics because each goal behavior is different. OAHG informs the user of the most optimal action for the current task. However, if the guidance is in conflict with the user's intention (i.e., a disagreement occurs), it will lead to an undesired physical interaction between the user and the robot, which can induce discomfort and frustration for the user~\cite{dragan2013policy, hernandez2015synthesizing}. On the other hand, UPHG supports comfortable movements of the users by reducing trajectory mismatch with the robot~\cite{de2015effect}, but has a limitation in that it cannot present more optimal movements, although they may exist. In this context, this study aims to answer the following questions, which have important implications for pHRI design, ``What is the difference between OAHG and UPHG in terms of user acceptance?'' and ``Is it possible to design a better HG by combining OAHG and UPHG?''

To impartially compare OAHG and UPHG, we need to implement each type of HG to achieve its best performance for a given application. Consequently, we present the following implementation methods to improve the performance of each HG based on deep learning approaches, which are attracting attention in recent HG studies~\cite{scobee2018haptic, wang2019comfort, broad2020data}. First, a deep reinforcement learning algorithm is used to train an optimal action model (for OAHG) to learn the optimal policy (i.e., the optimal way of behaving from a specific state during a task) through self-play between AI agents~\cite{bansal2018emergent} (the upper flow in Figure~\ref{fig:01}(b)). Second, we train a user prediction model (for UPHG) in a supervised manner with multiple user's behavior data. To deal with individual differences between human operators, we apply a meta-learning approach to enable to adapt the model parameters according to the current user (the lower flow in Figure~\ref{fig:01}(b)). Third, both the optimal action model and the user prediction model are designed to infer the model uncertainty from their outputs. By making HGs consider their model uncertainty at every timestep, we aim to prevent the decrease in HG performance because of an inaccurate model.

Another essential purpose of our study is to explore the possibility of complementing each HG type. If OAHG and UPHG are in a trade-off relationship with each other, it can be expected to produce HG that achieves optimal performance in terms of objective and subjective metrics by properly harmonizing the two HG types. In particular, we focus on the fact that the disagreement between the guiding force and the user intention reduces the OAHG performance ~\cite{dragan2013policy, hernandez2015synthesizing}. Therefore, we devised a method to implement the combined haptic guidance (CombHG) utilizing the similarity of guiding forces generated by the two HG types every timestep to minimize the disagreement. With this similarity-based method, the combined guiding force is adjusted according to the difference between the two guiding force directions.

We conducted a user experiment with 20 participants to verify how each type of HG has different effects on objective and subjective evaluations. We built a task environment for confronting an AI agent in an \textit{air hockey} game that can be operated and assisted through a haptic device, as shown in Figure~\ref{fig:01}(a). Each participant played the air hockey game with the HGs we implemented, and subjective evaluations including user interviews were conducted after each task. Our user experiment results indicate the following important points. First, all types of HG---OAHG, UPHG, and CombHG---led to a significant improvement in objective user performance compared to the case where the user did not receive any HG, but there was no significant difference in objective performance between the three HG types. Second, UPHG and CombHG elicited a significantly higher score in subjective metrics, such as perceived naturalness and comfort, than OAHG. Finally, CombHG significantly lowered the disagreement between the user intention and HG compared to the cases of OAHG and UPHG, without reducing the objective and subjective scores.

Overall, this paper has following three key contributions:
\begin{itemize}
  \item We present deep learning-based approach to implement OAHG and UPHG to achieve their best performance, applying a self-play-based reinforcement learning framework for OAHG and a meta-learning framework for UPHG. In particular, we propose and verify two novel implementation methods---uncertainty-based thresholding and user adapt\-ation---to improve the HG performance.
  \item We propose and verify a combined approach (CombHG) of OAHG and UPHG that can complement each HG type, utilizing our similarity-based combination method. To the authors' knowledge, this is the first attempt to combine OAHG and UPHG to achieve better performance.
  \item We experimentally compare OAHG, UPHG, and CombHG in terms of objective and subjective metrics through a user study and interview (n=20).
\end{itemize}

\section{Related Work}
\subsection{Haptic Guidance}
Early HG studies were based on control assistance methods in virtual or teleoperation environments, such as a virtual fixture~\cite{rosenberg1993virtual} and an artificial force field~\cite{xiao1998navigation}, which help users accurately move toward goals and prevent access to dangerous areas. Robotic devices enabled the delivery of haptic feedback generated by the virtual fixture or artificial force field techniques to users, allowing them to perform the tasks with improved stability~\cite{bettini2002vision, lam2009artificial}.  Over the past few decades, the implementation of HG has made significant progress, and it has been embedded in a variety of forms, including haptic devices~\cite{perez2016using, de2015effect, bluteau2008haptic, zahedi2017gesture}, sleeve devices~\cite{chen2016motion, goto2018artificial}, pen devices~\cite{kianzad2020phasking}, and steering wheels~\cite{mulder2008effect, wang2017effect, crespo2008haptic, hosseini2016predictive, tada2016simultaneous}. Accordingly, the range of HG applications has also expanded, such as in surgical assistance~\cite{zahedi2017gesture, power2015cooperative, hong2016haptic}, driving assistance~\cite{mulder2008effect, wang2017effect,hosseini2016predictive, tada2016simultaneous}, teleoperation of robots~\cite{perez2016using, selvaggio2019passive} and UAVs~\cite{lam2009artificial, smisek2016neuromuscular}, and desktop computer interfaces~\cite{dennerlein2000force, kuber2007towards}.

A majority of the previous studies have reported positive effects of HG, such as improvement in task performance and user comfort, and reduction of user workload. Nevertheless, some elements of HG that hamper usability remain. As the most representative example, a conflict between the user and the HG can lead to a temporary increase in the force exerted by the user~\cite{mulder2012sharing, boink2014understanding}, or even a decrease in performance~\cite{passenberg2013exploring}. In addition, if the interference of HG is excessive for a user, then the user requires more physical effort, which leads to a deteriorated user evaluation (e.g., low comfort and controllability)~\cite{mars2014analysis}. A detailed analysis of the factors affecting HG will bring useful implications in designing a user-friendly HG. In this study, we aim to determine what users expect (and not expect) from HG through a comparison between OAHG and UPHG, which was not sufficiently covered in previous studies. Furthermore, we attempt, for the first time, to optimally combine OAHG and UPHG to achieve better performance based on the understanding of the factors that influence HG performance.

\subsection{Implementing Optimal Action-based HG}
The most straightforward way of implementing OAHG is to set a reference trajectory for performing a task and deliver a continuous guiding force so that the user does not leave the trajectory. In the context of a steering task such as when driving, a number of studies have implemented OAHG in the form of a guiding force directed toward the center of the path~\cite{forsyth2005predictive, mulder2008effect, wang2017effect, crespo2008haptic, hosseini2016predictive}. As another example, in a backward parking situation, Tada \textit{et al.}~\cite{tada2016simultaneous} implemented OAHG by utilizing a Bezier curve between a start point and a target parking point as a reference trajectory. Meanwhile, when the reference trajectory could not be clearly defined in advance, a demonstration by skilled experts also served as the reference trajectory, for example, in the case of a handwriting task~\cite{bluteau2008haptic, teranishi2017effects} and a peg-in-hole task~\cite{perez2016using}.

Recently, attempts have been made to apply reinforcement learning techniques in HG implementation to train an optimal policy for a task. Scobee \textit{et al.}~\cite{scobee2018haptic} attempted to determine the underlying value function of each observation state of a steering task from the movement of an expert operator through the inverse reinforcement learning method, and developed OAHG based on the trained value function. Meanwhile, the deep Q-network (DQN) was applied in~\cite{wang2019comfort} to train HG that minimizes the magnitude of the steering wheel angle during a steering task. These previous reinforcement learning-based HGs showed promising results, but had a limitation in that they cannot provide a perfectly optimal action because they were trained by data from very few (one or two) human operators. In this study, we trained an optimal action model for implementing OAHG by using a latest reinforcement learning framework based on self-play between AI agents only~\cite{bansal2018emergent}.

\subsection{Implementing User Prediction-based HG}
Research on UPHG emerged later than research on OAHG and has received less attention. De Jonge \textit{et al.}~\cite{de2015effect} presented UPHG according to a reference trajectory adapted to each user based on his/her previous trial in a steering task, whereas OAHG utilized a fixed reference trajectory (i.e., center of the path). Their personalized UPHG demonstrated the positive effect of reducing the conflict between the human operator and HG. Meanwhile, data-driven approaches have also been used to model a user's movement. Hern{\'a}ndez \textit{et al.}~\cite{hernandez2015synthesizing} stochastically modeled each user's individual movement based on a hidden Markov model (HMM) in a task of moving a virtual object to a target position avoiding obstacles, and presented UPHG to assist the user's movement based on the user model. Previous studies such as~\cite{power2015cooperative} and~\cite{zahedi2017gesture} also presented HG based on a user movement model using HMMs in a surgical task environment.

In this study, we implement a user model for UPHG based on deep neural networks. Such an approach has recently been highlighted for user modeling in HCI and HRI fields~\cite{schmerling2018multimodal, moon2019prediction, pfau2020bot, moon2020sample}. To apply UPHG tailored for an individual user, it is necessary to learn on each individual's data. However, collecting enough data to train the neural network-based model from scratch for every new user is very inefficient, particularly, when targeting numerous users. To solve this challenging problem, we apply a meta-learning approach, enabling the trained model to adapt the model parameters to a new user based on his/her short trial data, which is the first attempt in UPHG implementation.

\section{Haptic Guidance Design}
\subsection{Target Task Environment}
\begin{figure}[t]
  \centering
  \includegraphics[width=\linewidth]{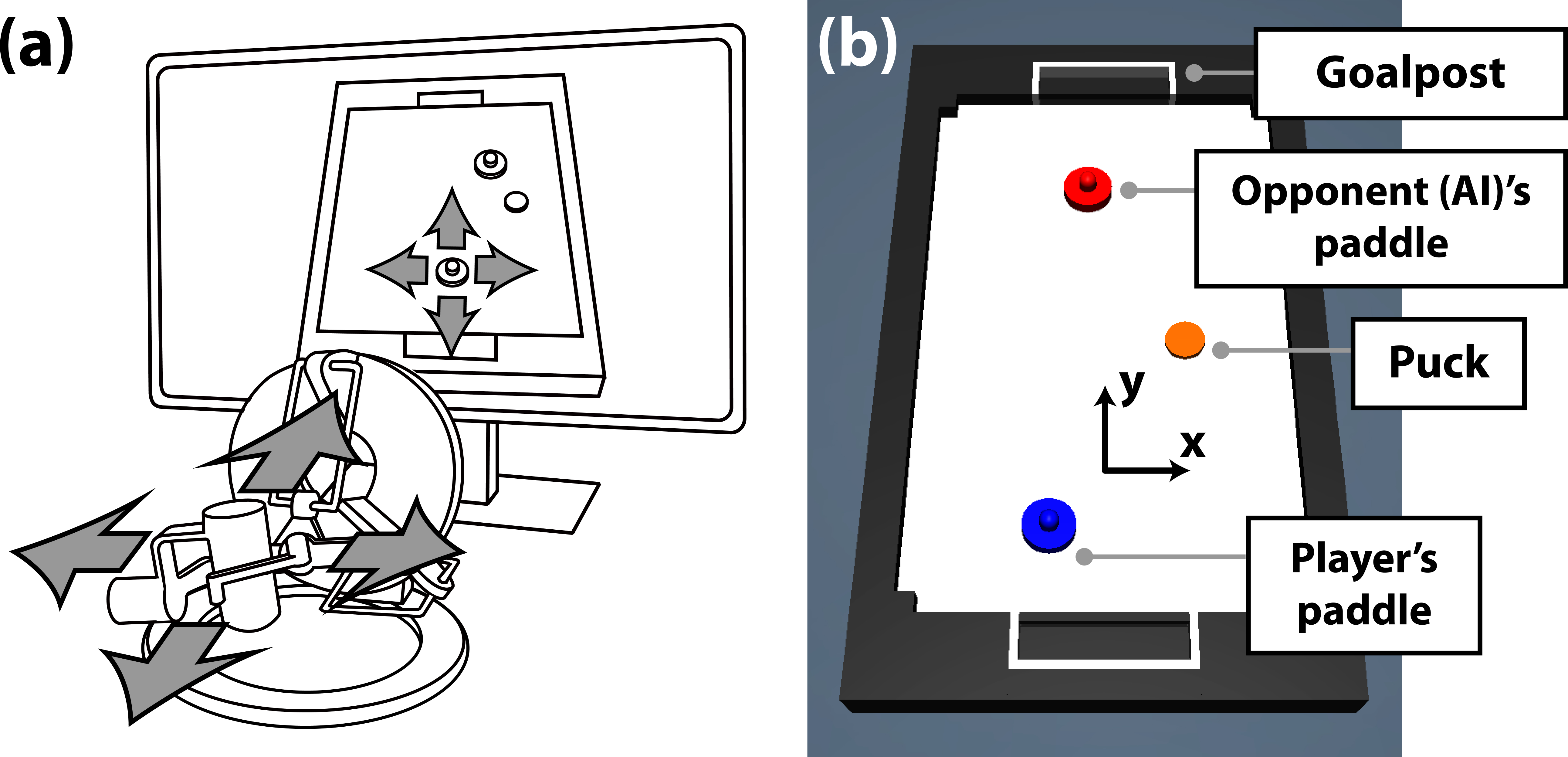}
  \caption{The air hockey game environment we developed. (a) Example of control based on a haptic device. (b) In-game screen and composition of our air hockey game.}
  \label{fig:02}
\end{figure}

As a target task for the experimental investigation of various types of HG, we developed a virtual \textit{air hockey} game environment that can be controlled using a haptic device (Figure~\ref{fig:02}(a)). In our air hockey game environment, a player competes with an opponent AI on a slippery surface by moving his/her \textit{paddle}, which is used to hit a \textit{puck} (Figure~\ref{fig:02}(b)). The player's successful task execution is judged by how many times he/she wins against the opponent AI. The player wins a round if he/she succeeds in putting a puck into the opposing goalpost, whereas he/she loses the round if the opponent puts a puck in the player's goalpost. Each paddle is set to move only within its own side (i.e., above or below the half-line). The player's paddle is controlled by a two-dimensional action vector, which corresponds to the desired paddle location on the xy coordinates with the midpoint of the player's side as the origin (Figure~\ref{fig:02}(b)). The action vector has one-to-one correspondence with the position of the end effector of the haptic device. For intuitive control of the player, the end effector is also set to move only on the 2D plane, and the moving directions of the end effector and the paddle are matched (Figure~\ref{fig:02}(a)).

A video game environment is suitable for comparing OAHG and UPHG in that each player can adopt various strategies according to his/her preference~\cite{iacovides2014learning, poeller2018let} while having a common winning formula. A player's basic strategy to win the air hockey game is to accurately \textit{smash} the puck toward the opponent's goal and \textit{defend} against the puck from the opponent heading inside the player's goal. However, in a detailed process, players can have several choices. A player can smash along a path that goes directly toward the goal, or he/she can choose a path through one or two reflections using the wall, avoiding the opponent's paddle. In addition, in a defensive situation, that is, when the opponent has the puck, a player can press the opponent near the half-line to narrow the angle of attack or wait for the opponent's attack right in front of the goal. We expect UPHG to predict the player's choices in advance and assist him/her with the corresponding action, while OAHG informs him/her of the most optimal action he/she can choose.

\subsection{Optimal Action Model and User Prediction Model}
We implemented the following two models based on deep neural networks: 1) an optimal action model that outputs the most optimal action that a user can take and 2) a user prediction model that outputs the expected action to be selected by the current user. Because both models use the same type of input (i.e., current game state) and output (i.e., target action vector), we designed them to have the same structure but with different hyperparameters (i.e., depth and size of the hidden layer) that are fine-tuned for each model's training. Each model was trained through different learning approaches based on reinforcement learning and meta-learning, respectively.

\subsubsection{Model Architecture}
\begin{figure}[t]
  \centering
  \includegraphics[width=0.7\linewidth]{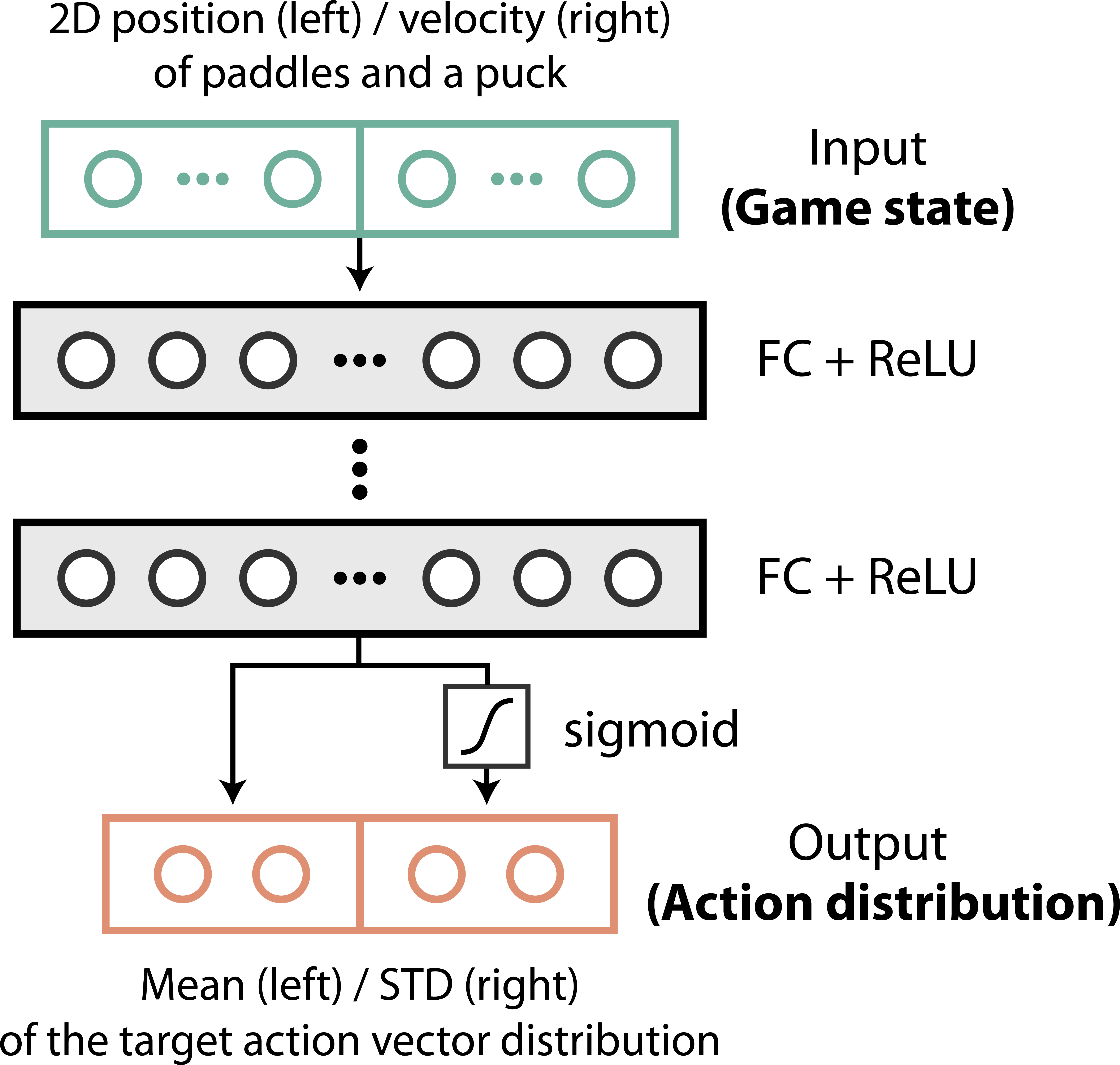}
  \caption{Overview of the model structure applied to both the optimal action model and the user prediction model.}
  \label{fig:03}
\end{figure}

Figure~\ref{fig:03} shows the structure of our model. The air hockey game state---the two-dimensional position and velocity vectors of two paddles and a puck (total, size of 12)---is mapped to the two-dimensional distribution data, which are the mean and standard deviation (STD) of the distribution of each xy coordinate of the action vector (total, size of 4). The output distribution represents the distribution of the target action vector used to generate the guiding force (i.e., optimal action for OAHG or predicted action for UPHG). Notably, we designed the model to output the mean and STD of the distribution rather than a single action value. In our HG implementation, we utilize the mean of the distribution as a reference action to guide the users, and also consider the model uncertainty which can be inferred from the STD of the distribution. Fully connected (FC) hidden layers with rectified linear unit (ReLU) activation functions are present between the input and output nodes. We used grid search to determine the optimal hyperparameters that exhibit the best learning performance in each training method; therefore, the depth and size of the hidden layers are different for the optimal action model and user prediction model: \(2\times64\) and \(4\times80\), respectively. The output STD goes through an additional sigmoid activation function.

\subsubsection{Optimal Action Model Training}
Reinforcement learning techniques have succeeded in acquiring an optimal policy that surpasses a human player in simulated video game environments~\cite{mnih2015human, silver2017mastering}. To train the optimal action model for our air hockey task, which is based on the competition between two players, we applied a latest learning framework based on self-play between two AI agents~\cite{bansal2018emergent}. In this framework, a training agent of the first generation initially confronts an opponent with random movements and updates the action model for increasing the expected reward. As the learning progresses, the training agent of the next generation competes with the agents of the previous generations, which also grow gradually, therefore the optimal action model can be progressively developed without human operator intervention. Meanwhile, the training agent updates the model using the sampled action from the action distribution output of the model. Therefore, when the specific action is confidently more optimal than other actions, the model will train to output a lower STD value, that is, the STD reflects the model uncertainty at current state.

For implementation details, trust region policy optimization (TRPO)~\cite{schulman2015trust} was used as the learning algorithm that updates the policy of the training agent, because it internally had the best learning performance for our air hockey task among the latest algorithms such as proximal policy optimization (PPO)~\cite{schulman2017proximal} and soft actor-critic (SAC)~\cite{haarnoja2018soft}.  Self-play-based procedural learning was performed over 100 generations. Each generation was trained with a simulation of 200K timesteps, and batch updates were applied every 5K timesteps.

\subsubsection{User Prediction Model Training}
In utilizing the user prediction model, we mainly focus on how to train the model to adapt to different users. Inspired by the fact that the existing meta-learning frameworks train models to quickly learn to perform new unseen tasks using only a few datapoints of each new task, we applied a meta-learning approach to train the user prediction model for HG to quickly adapt to different users. We demonstrated in our prior study~\cite{moon2020dynamic} that model-agnostic meta-learning (MAML)~\cite{finn2017model}, a recent mainstream meta-learning algorithm, can be used to train a model that mimics user behavior for dynamic difficulty adjustment and effectively adapts to a new user with his/her minimal demonstration data. In the current study, we present a \textit{user adaptation} (UA) method that trains the user prediction model based on the MAML algorithm and updates the model parameters for each user. The training process of our user prediction model is as follows. The training data consists of a set of task execution data \(D^U\) of different users \(U\), and \(D^U\) is divided into \(D_{demo}^U\) for model adaptation, and \(D_{valid}^U\) for meta-update. First, the algorithm updates the model parameter \(\theta\) by a single gradient descent that reduces \(L(\theta,D_{demo}^U)\), which is the model loss function for \(D_{demo}^U\). Therefore,
\begin{equation}
  \theta_{adapt}^U=\theta-\alpha\nabla_\theta L(\theta,D_{demo}^U),
  \label{eq:01}
\end{equation}
where \(\alpha\) denotes the inner learning rate. Subsequently, a meta-update is performed to reduce model loss for \(D_{valid}^U\) using the updated parameters. Overall, the objective of our meta-learning is as follows:
\begin{equation}
  min\sum_U L(\theta_{adapt}^U,D_{valid}^U).
  \label{eq:02}
\end{equation}
Through this two-fold backpropagation, the user prediction model eventually learns to rapidly adapt to unseen users. Meanwhile, we aim to train the model to output the STD of the predicted action distribution, which is impossible with the mean-squared error loss function commonly used in general regression learning. Inspired by a previous work~\cite{kendall2017what} that modeled uncertainty in deep learning models for computer vision tasks, we use the following loss function to train our user prediction model:
\begin{equation}
  L(\theta,D)=\frac{1}{N}\sum\nolimits_{(\mathbf{x}_i,\mathbf{y}_i)\in D}\frac{1}{2\|\bm{\hat{\sigma}}_i\|^2}\|\mathbf{y}_i-\mathbf{\hat{y}}_i\|^2+\frac{1}{2}\log\|\bm{\hat{\sigma}}_i\|^2,
  \label{eq:03}
\end{equation}
where \(N\) denotes the size of the training data \(D\) (i.e., \(D_{demo}^U\) or \(D_{valid}^U\)), subscript \(i\) indicates the \(i\)-th sample from \(D\), \(\mathbf{y}_i\) is an actual action taken by a player at a game state \(\mathbf{x}_i\), and \(\mathbf{\hat{y}}_i\) and \(\bm{\hat{\sigma}}_i\) are the predicted mean and STD output by feeding \(\mathbf{x}_i\) into the model with parameter \(\theta\). With this loss function, the model can be trained to output the STD that implies the uncertainty of the predicted action; that is, the lower the STD values, the more confident the model is about the prediction.

For the user prediction model training, we collected data from nine participants aged between 22--29 (mean=25.33, STD=2.16) while performing the air hockey task without haptic guidance. Each participant was provided sufficient practice time to be familiar with the task prior to data collection. A total of 360K timesteps of recorded behavioral data were employed to train the user prediction model. During our meta-learning process, the adaptation was conducted with an inner learning rate of 0.1, and the meta-update was conducted using an Adam optimizer with a learning rate of 0.001. The batch size of \(D_{demo}^U\) and \(D_{valid}^U\) was 1K timesteps, and the entire training was conducted for 200 epochs.

To verify the practical effectiveness of the UA method, in a subsequent user study, we experimentally compared the performance of two UPHGs based on models trained with and without UA. For the UPHG implementation without using the UA method, we trained another user prediction model in a typical supervised manner, excluding the meta-learning approach, employing the same dataset used in the training with UA. In this case, the model trained without UA will predict generalized user movement because it is trained to use fixed network parameters for various user behaviors. An Adam optimizer with a learning rate of 0.001 was used and the batch size was 1K timesteps. The entire training was conducted for 100 epochs.

\subsection{Generating HGs Based on Trained Models}
The trained models output the target action for assisting the user's task. A straightforward and reliable approach to generate HG is to implement a spring force toward the target action so that users can match their actions accordingly~\cite{forsyth2005predictive, mulder2008effect, de2015effect, scobee2018haptic}. Meanwhile, one of the important implications from the previous studies on HG implementation is that a human operator cannot respond to HG immediately. Accordingly, a lookahead method~\cite{forsyth2005predictive} where HG should be proactively applied based on a slight future game state, considering the reaction time of a human, has been applied as a common technique. Following the method described in~\cite{forsyth2005predictive}, we used the anticipated game state \(\mathbf{x}_{lookahead}\), which is obtained by virtually moving the puck and paddles with the current velocity for a short time \(T_{lookahead}\), instead of the current game state \(\mathbf{x}\) as the model input. Therefore, the guiding force of our OAHG and UPHG is generated by the following equation:
\begin{equation}
  \mathbf{F}_{HG}=clip(-K(\mathbf{u}-\mathbf{\hat{y}})),
  \label{eq:04}
\end{equation}
where \(K\) is the stiffness gain, \(\mathbf{u}\) is the user's current action vector, and \(\mathbf{\hat{y}}\) is the mean of the target action vector distribution obtained by feeding \(\mathbf{x}_{lookahead}\) into the model. A clipping function was applied to make the guiding force bounded, because momentary excessive force may induce safety problems to the user.

\begin{figure}[b]
  \centering
  \includegraphics[width=\linewidth]{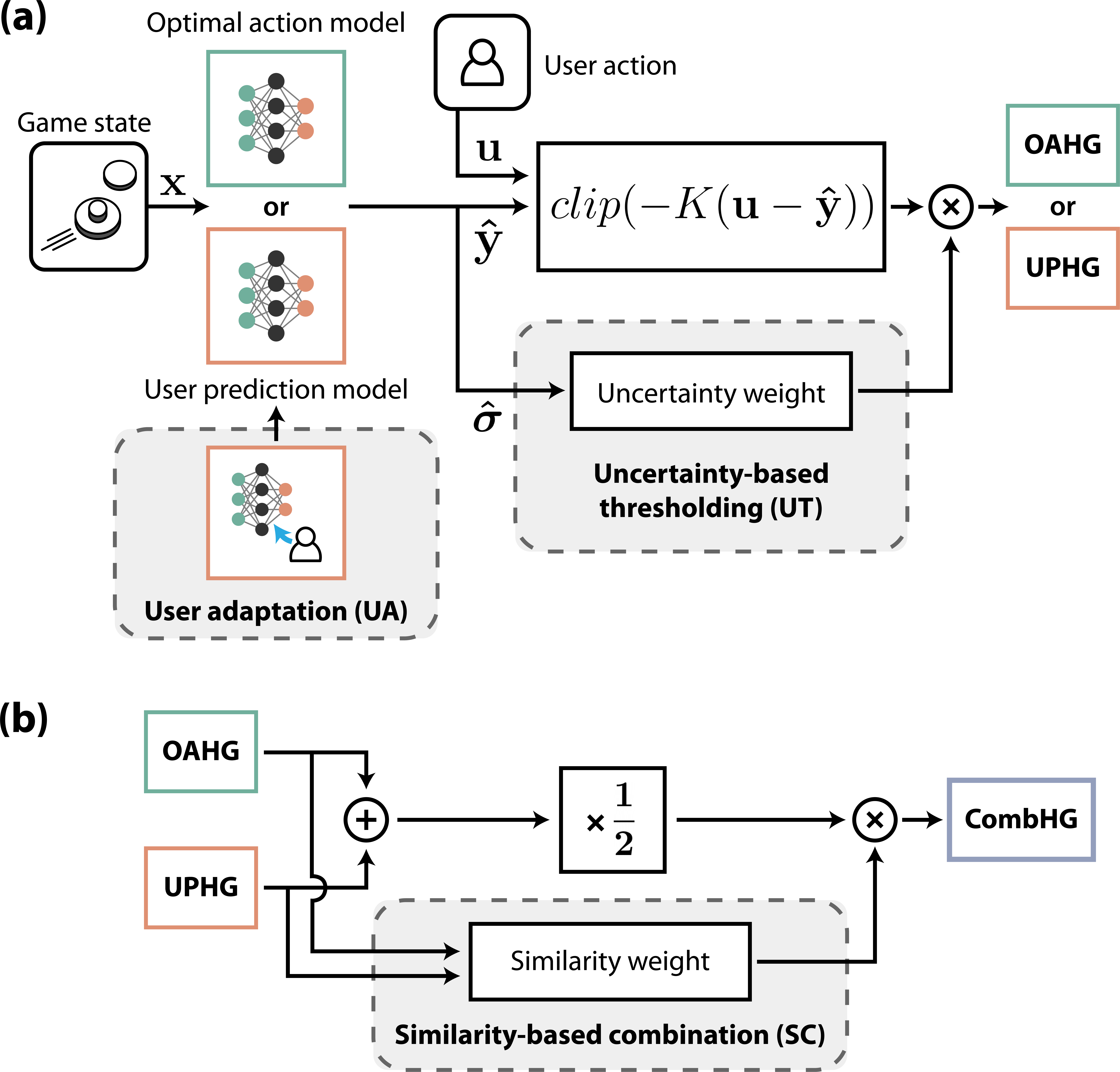}
  \caption{The process of implementing (a) optimal action-based haptic guidance (OAHG) and user prediction-based haptic guidance (UPHG), and (b) combined haptic guidance (CombHG). The three proposed methods (i.e., UT, UA, and SC), whose effectiveness is investigated through a subsequent user experiment, are highlighted in gray shades.}
  \label{fig:04}
\end{figure}

\begin{table*}[t]
  \caption{Proposed methods applied to our HG implementation.}
  \label{tab:01}
  \begin{tabular}{clccc}
    \toprule
    \textbf{Abbrev}& \textbf{Full Form}& \textbf{OAHG}& \textbf{UPHG}& \textbf{CombHG}\\
    \midrule
    \textbf{UT}& Uncertainty-based thresholding& \checkmark& \checkmark& \checkmark\\
    \textbf{UA}& User adaptation& & \checkmark& \checkmark\\
    \textbf{SC}& Similarity-based combination& & & \checkmark\\
    \bottomrule
    \multicolumn{5}{l}{\footnotesize{\checkmark A checkmark indicates whether it was used for the corresponding HG type.}}
  \end{tabular}
\end{table*}

Additionally, we propose an \textit{uncertainty-based thresholding} (UT) method that adjusts the magnitude of HG by considering model uncertainty. The importance of model uncertainty has been recognized in several previous pHRI studies~\cite{hernandez2015synthesizing, grau2013effect, kanazawa2019adaptive}, as the results from inaccurate models of robot or human movements can lead to human discomfort or safety issues during human–robot collaborations. We infer the model uncertainty at a specific state from the STD of the target action; that is, if the output STD is large, the model at current state is judged to be highly uncertain. Using this, we defined the uncertainty weight according to the squared magnitude of the STD as follows:
\begin{equation}
  W_{unc}=
  \begin{cases}
    1, & \text{if } \|\bm{\hat{\sigma}}\|^2<T_{low} \\
    \frac{T_{high}-\|\bm{\hat{\sigma}}\|^2}{T_{high}-T_{low}}, & \text{if } T_{low}\leq\|\bm{\hat{\sigma}}\|^2<T_{high} \\
    0, & \text{otherwise,}
  \end{cases}
  \label{eq:05}
\end{equation}
where \(\bm{\hat{\sigma}}\) is the output STD, and \(T_{low}\) and \(T_{high}\) are threshold values. With the UT method, calculated \(\mathbf{F}_{OAHG}\) or \(\mathbf{F}_{UPHG}\) from (\ref{eq:04}) is multiplied by the uncertainty weight, which ranges from 0 to 1, and then provided to users. In other words, only when the STD is less than a certain level (i.e., \(T_{high}\)), we determine the HG to be confident in its purpose and can assist the users. Figure~\ref{fig:04}(a) summarizes the process of generating OAHG and UPHG, including the UT method.

A straightforward approach to implement CombHG is to use the average of two guiding force vectors from OAHG and UPHG, therefore make the user to be assisted by both HG types simultaneously. This approach is intuitive in that the combined HG guides the user to the midpoint between the two target actions to be guided by OAHG and UPHG. Accordingly, we implement the simplest CombHG as follows:
\begin{equation}
  \mathbf{F}_{CombHG}=(\mathbf{F}_{OAHG}+\mathbf{F}_{UPHG})/2.
  \label{eq:06}
\end{equation}

However, simply taking the average of two guiding forces may worsen the results, guiding the user to an unintended third direction if the two HGs have different directions. To solve this problem, we propose a conservative combination method, namely \textit{similarity-based combination} (SC), that considers the similarity of two HGs based on the angle between the two guiding forces and reduces the magnitude of HG when the similarity is low. The main purpose of the SC method is to provide an appropriate guiding force only when the target action is optimal and meets the user's intention, that is, when the directions of the two HGs match. To do this, we define the similarity weight as follows:
\begin{equation}
  W_{sim}=\cos^2(\phi/2),
  \label{eq:07}
\end{equation}
where \(\phi\) is the angle between the two guiding forces according to OAHG and UPHG. The similarity weight ranges from 0 to 1, which corresponds to the angle \(\phi\) from \(\pi\) (opposite direction) to 0 (matched direction). With the SC method, \(\mathbf{F}_{CombHG}\) from (\ref{eq:06}) is provided to the user after multiplying by the similarity weight. Figure~\ref{fig:04}(b) shows the implementation of CombHG using OAHG and UPHG, including the SC method.

\section{User Study Design}
\subsection{Research Questions}
The objective of this research is divided into two main parts: a presentation of the implementation methods for three types of HG, that is, OAHG, UPHG, and CombHG, and an experimental investigation into how each type of HG differs in the context of user acceptance. We implemented the HGs by generating a guiding force based on the real-time output of the optimal action model and the user prediction model we trained. To achieve better performance of HG, we applied the three proposed methods (i.e., UT, UA, and SC), according to the HG type specified in Table~\ref{tab:01}. To aid future HG studies, this user study first aims to experimentally confirm whether the methods actually lead to a HG performance improvement. Subsequently, by using HGs to which all the applicable methods are employed, we determine whether each HG assists users when compared to a no HG (NHG) condition. Further, we investigate how the three types of HG assist users differently in terms of objective and subjective evaluations, which is our primary research objective. Therefore, we formulate the following four research questions:

\begin{itemize}
 \item \textbf{RQ1:} Do the UT, UA, and SC methods that we propose and apply to the HG implementation contribute to an improvement in HG performance?
 \item \textbf{RQ2:} Do OAHG, UPHG, and CombHG improve users' task performance when compared to NHG?
 \item \textbf{RQ3:} What differences do OAHG and UPHG have in users' objective and subjective evaluations?
 \item \textbf{RQ4:} Can CombHG, which integrates OAHG and UPHG, complement each HG or provide better effects in users' objective and subjective evaluations?
\end{itemize}

\subsection{Experimental Method}
We conducted an indoor laboratory experiment to measure the actual assisting performance of the implemented HGs. We recruited 20 participants (4 females and 16 males) aged between 21--30 (mean=\allowbreak25.58, STD=2.09) for this user study. All participants were right-handed, and none of them reported perception defects in their vision and touch.

\begin{figure}[t]
  \centering
  \includegraphics[width=0.7\linewidth]{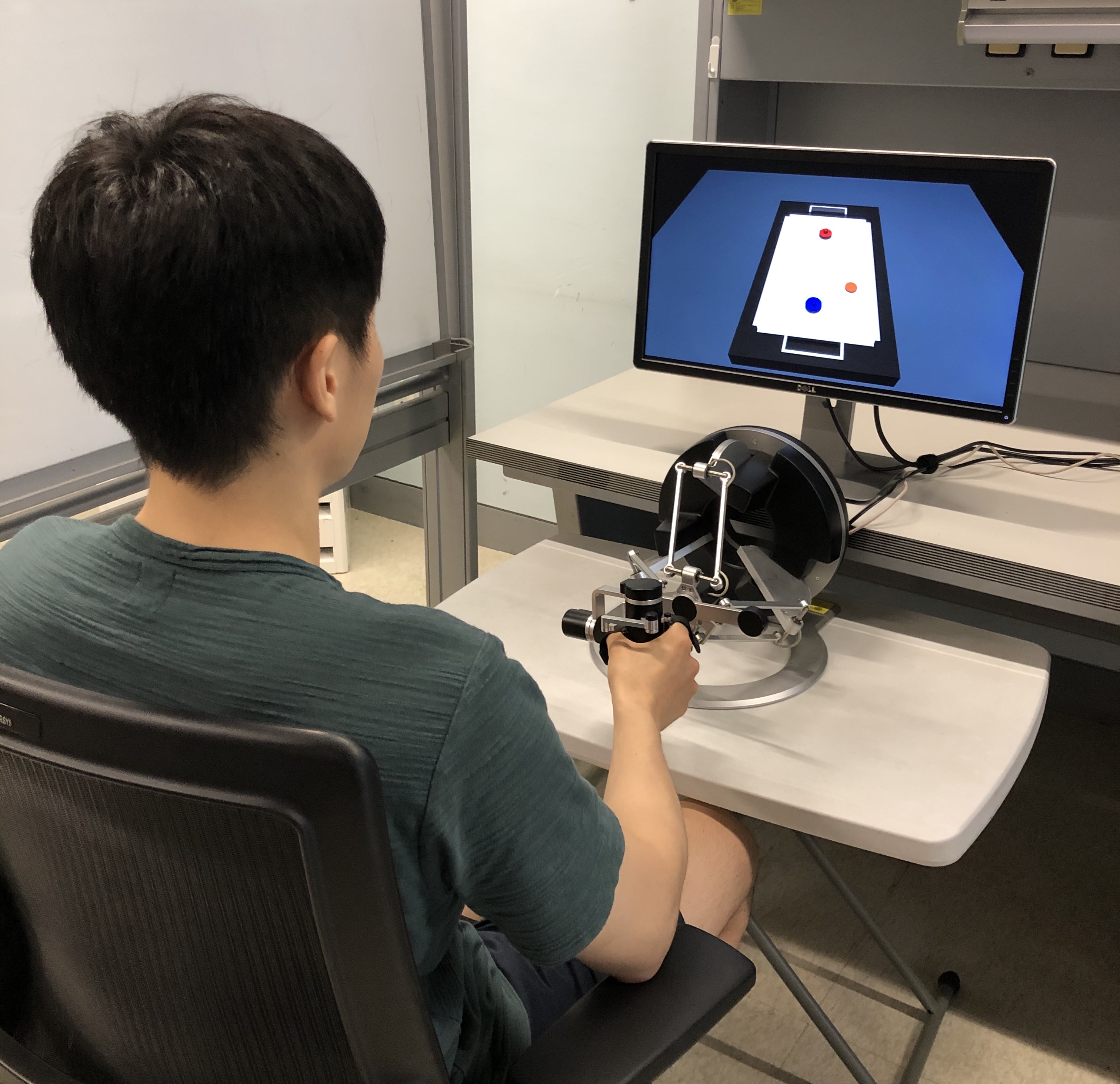}
  \caption{Experimental setup of our user study. A participant performs a virtual air hockey task with a haptic device.}
  \label{fig:05}
\end{figure}

Figure~\ref{fig:05} presents the experimental setup. The air hockey task environment was composed of a haptic interface device (Omega.7, Force Dimension) and a 24-inch monitor, which were connected to a PC. Each participant was instructed to sit down and perform the air hockey task by holding the end effector of the haptic device with his/her dominant hand.

We selected HGs under the following nine conditions for the user experiment for verifying the effectiveness of the HG implementation methods (RQ1) and comparing the effect of each HG type (RQ2--RQ4):

\renewcommand{\labelenumi}{(\alph{enumi})}
\begin{enumerate}
  \item \textbf{NHG:} no haptic guidance.
  \item \textbf{OAHG (vanilla):} OAHG from Equation (\ref{eq:04}) with no additional method.
  \item \textbf{OAHG (UT):} OAHG applying the uncertainty-based thresholding in addition to OAHG (vanilla).
  \item \textbf{UPHG (vanilla):} UPHG from Equation (\ref{eq:04}) using a model not trained with the meta-learning algorithm, and therefore without the user adaptation.
  \item \textbf{UPHG (UA):} UPHG from Equation (\ref{eq:04}) using a model trained with the meta-learning algorithm (i.e., UA applied).
  \item \textbf{UPHG (UT+UA):} UPHG applying the uncertainty-based thresholding in addition to UPHG (UA).
  \item \textbf{CombHG (UA):} CombHG from Equation (\ref{eq:06}) based on OAHG (vanilla) and UPHG (UA). Note that because the basic concept of UPHG is assisting a user according to personalized prediction, we use UPHG (UA) instead of UPHG (vanilla) to implement the underlying CombHG.
  \item \textbf{CombHG (UA+SC):} CombHG applying the similarity-based combination on OAHG (vanilla) and UPHG (UA).
  \item \textbf{CombHG (UT+UA+SC):} CombHG applying the similarity-based combination on OAHG (UT) and UPHG (UT+UA).
\end{enumerate}

Our experimental procedure is as follows. First, all participants were informed that they would experience nine different HGs, and interviews would be conducted about their impressions received from each HG, immediately after experiencing each HG. Second, each participant was provided with as much practice time as he/she wanted to be familiar with the task environment. Third, for the purpose of updating the user prediction model parameters (i.e., UA), each participant performed one game in the NHG condition, which was not reflected in the evaluation. One game basically consists of seven rounds (one round ends when either a player or an AI scores a goal), but if it ends earlier than the minimum play time that we set (two minutes per task), up to three additional rounds proceed to secure more data from the player. After completing UA with the recorded data, that is, updating the parameters of the user prediction model through Equation (\ref{eq:01}), the participant sequentially performed all nine HG conditions in a random order with counterbalancing. All participants performed one game for each HG condition, and before the start of each game, they were allowed 30 seconds to adapt to the given HG, which was not reflected in the evaluation. Following the end of each game, participants assessed the subjective scores for the HG they had just been assisted with, and a short interview with a supervising researcher was conducted. All participants were able to rest as much as they wanted between each game. The total duration of the experiment was between 1--1.5 hours, depending on the participants.

\subsection{Measured Variables and Metrics}
We measured the assisting performance of each HG condition in terms of objective metrics automatically measured by the system and subjective metrics based on participants' evaluations. Four objective metrics were used: win rate, mean smash speed, and defense rate, which indicate the degree of high task performance; and mean disagreement, which indicates the degree of high conflict between a user and HG. The metrics are defined as follows:

\begin{itemize}
 \item \textbf{Win rate:} the ratio of the participant winning the opponent in one game (7--10 rounds).
 \item \textbf{Mean smash speed:} the average speed of the puck hit by the participant over the opponent's side (i.e., smashed). Note that, the speed was measured in relative figures because the air hockey task was built in a virtual environment without specific units.
 \item \textbf{Defense rate:} the proportion of the pucks blocked by the participant among the pucks headed into the participant's goal.
 \item \textbf{Mean disagreement:} the average of the disagreement between the participant and HG, which is proposed in~\cite{hernandez2015synthesizing} and defined as follows (\(\bm{\Delta}\mathbf{u}\) denotes the user's action change during a timestep after receiving HG).
\end{itemize}
\begin{displaymath}
  disagreement=
  \begin{cases}
    -\frac{\mathbf{F}_{HG}^T\cdot\bm{\Delta}\mathbf{u}}{\|\bm{\Delta}\mathbf{u}\|}, & \text{if } \mathbf{F}_{HG}^T\cdot\bm{\Delta}\mathbf{u}<0 \\
    0, & \text{otherwise.}
  \end{cases}
\end{displaymath}

\begin{figure*}[t]
  \centering
  \includegraphics[width=\linewidth]{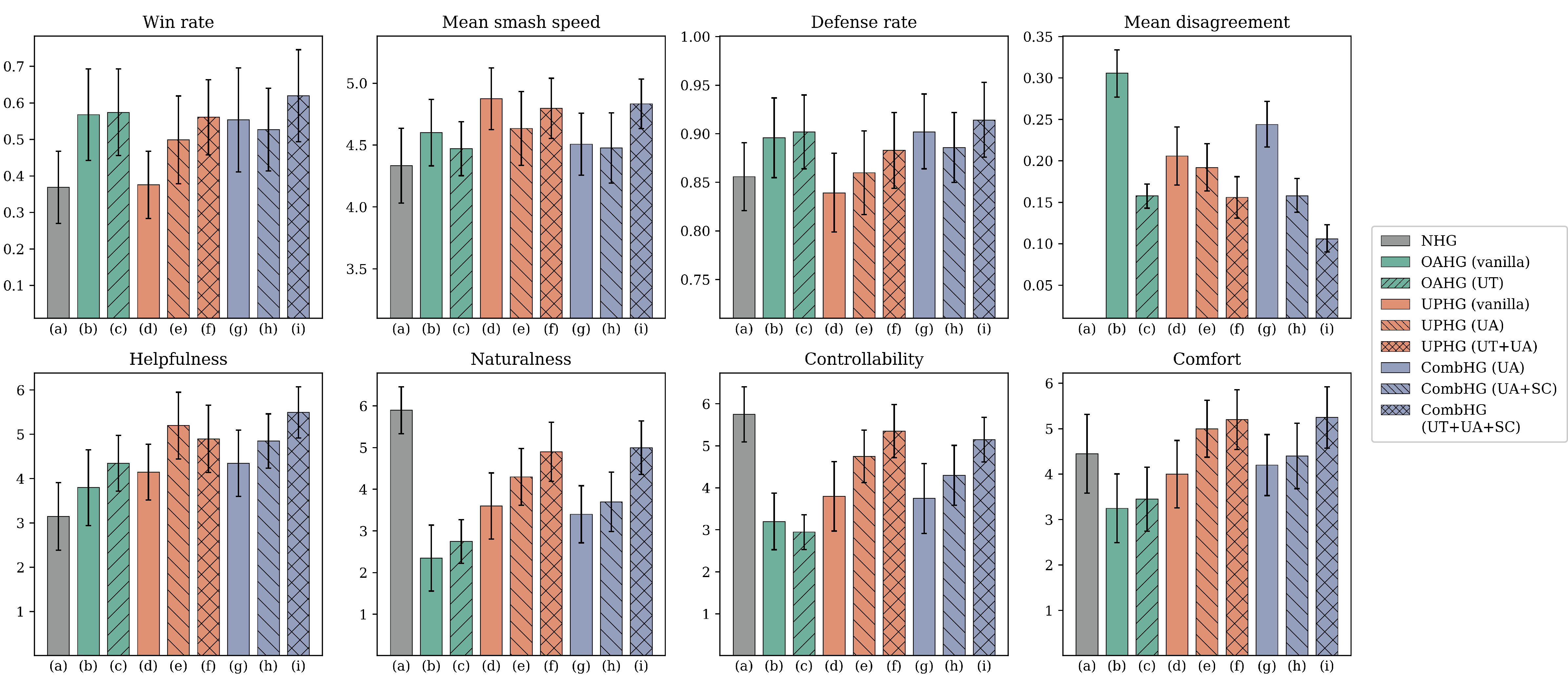}
  \caption{Experimental results of the objective (top) and subjective (bottom) metrics for the nine HG conditions (a)–-(i). The HG conditions of the same HG type are indicated by bars of the same color. Error bars represent 95\% confidence intervals.}
  \label{fig:06}
\end{figure*}

For the subjective evaluation, we selected the following four items and asked the participants to rate each HG condition in terms of the items on a 7-point Likert score:

\begin{itemize}
 \item \textbf{Helpfulness:} how much the participants felt the HG \textit{helped} them perform the task.
 \item \textbf{Naturalness:} how \textit{natural} the participants felt the assistance of the HG.
 \item \textbf{Controllability:} how well the participants felt they were \textit{able to control} the paddle under the HG.
 \item \textbf{Comfort:} how \textit{comfortable} the participants were with the assistance of the HG.
\end{itemize}

\section{Results}

Figure~\ref{fig:06} shows the objective and subjective evaluation results for each HG condition. To answer RQ1, we first investigated how assisting performance varies according to implementation methods within each HG type. Next, we examined how each HG type provides different assistance for the user (RQ2--RQ4), using the results of the HG conditions to which all applicable implementation methods were employed, that is, OAHG (UT), UPHG (UT+UA), and CombHG (UT+UA+SC).

\subsection{Comparison Within HG Types}
We first compared the evaluated results of the two conditions within the OAHG type according to the implementation methods (i.e., vanilla vs. UT). A paired t-test revealed that the use of the UT method in the OAHG type significantly reduces the participants' mean disagreement (\(t=12.143\), \(p<.001\)). On the other hand, no significant difference was found in the remaining seven metrics between the OAHG (vanilla) and OAHG (UT) conditions.

To compare the three conditions within the UPHG type (i.e., vanilla, UA, UT+UA), a repeated measures ANOVA with a Green\-house-Geisser correction was used. The analysis revealed that there were significant effects of the UPHG implementation methods on the following metrics: win rate (\(F_{2,38}=6.430\), \(p=.004\), \(\eta^2=.253\)), mean disagreement (\(F_{2,38}=19.856\), \(p<.001\), \(\eta^2=.511\)), helpfulness (\(F_{2,38}=3.422\), \(p=.044\), \(\eta^2=.153\)), naturalness (\(F_{2,38}=5.156\), \(p=.013\), \(\eta^2=.213\)), controllability (\(F_{2,38}=10.188\), \(p=.001\), \(\eta^2=.349\)), and comfort (\(F_{2,38}=7.012\), \(p=.006\), \(\eta^2=.270\)). However, there was no significant effect on the mean smash speed and defense rate. For the metrics with significant effects, we conducted post hoc tests with a Bonferroni correction to investigate a significant mean difference between the conditions. Table~\ref{tab:02} summarizes the comparisons of the conditions in which significant mean differences exist (i.e.,  \(p<.05\)). According to the analysis, the use of methods such as UA and UT contributed individually or together to improve objective evaluations (e.g., an increase in win rate and a decrease in mean disagreement) and subjective evaluations (e.g., an increase in helpfulness, naturalness, controllability and comfort) from participants.

\begin{table*}[t]
  \caption{Comparisons showing statistically significant differences between implementation methods within HG type.}
  \label{tab:02}
  \begin{tabular}{clccclcc}
    \toprule
    \textbf{Type}& \textbf{Metrics}& \textbf{Comparison}& \textbf{\(p\)}& \textbf{Type}& \textbf{Metrics}& \textbf{Comparison}& \textbf{\(p\)}\\
    \midrule
    \textbf{OAHG}& Mean disagreement& (b) \(>\) (c)& .000&
    \textbf{CombHG}& Mean smash speed& (i) \(>\) (g)& .003\\
    \textbf{UPHG}& Win rate& (f) \(>\) (d)& .009&
     & & (i) \(>\) (h)& .028\\
     & Mean disagreement& (d) \(>\) (f)& .000&
     & Mean disagreement& (g) \(>\) (h)& .000\\
     & & (e) \(>\) (f)& .000&
     & & (g) \(>\) (i)& .000\\
     & Helpfulness& (e) \(>\) (d)& .041&
     & & (h) \(>\) (i)& .000\\
     & Naturalness& (f) \(>\) (d)& .032&
     & Helpfulness& (i) \(>\) (g)& .023\\
     & Controllability& (e) \(>\) (d)& .028&
     & Naturalness& (i) \(>\) (g)& .001\\
     & & (f) \(>\) (d)& .004&
     & & (i) \(>\) (h)& .043\\
     & Comfort& (f) \(>\) (d)& .001&
     & Controllability& (i) \(>\) (g)& .009\\
     & & & &
     & Comfort& (i) \(>\) (g)& .023\\
    \bottomrule
  \end{tabular}
\end{table*}

We also conducted a repeated measures ANOVA with a Green\-house-Geisser correction to compare the results of the conditions representing the three implementation methods within the CombHG type (i.e., UA, UA+SC, and UT+UA+SC). There were significant effects of the implementation methods on the following metrics: mean smash speed (\(F_{2,38}=5.733\), \(p=.011\), \(\eta^2=.232\)), mean disagreement (\(F_{2,38}=79.038\), \(p<.001\), \(\eta^2=.806\)), helpfulness (\(F_{2,38}=3.295\), \(p=.049\), \(\eta^2=.148\)), naturalness (\(F_{2,38}=7.883\), \(p=.002\), \(\eta^2=.293\)), controllability (\(F_{2,38}=5.607\), \(p=.008\), \(\eta^2=.228\)), and comfort (\(F_{2,38}=4.057\), \(p=.027\), \(\eta^2=.176\)). However, there was no significant effect on win rate and defense rate. Post hoc tests with a Bonferroni correction were also conducted, and comparisons between conditions with significant mean differences are summarized in Table~\ref{tab:02}. Similar to the analysis of the UPHG type, the post hoc analysis indicated that the methods we present for CombHG (i.e., UT and SC) contribute to improving the objective and subjective aspects of assisting performance for users.

\subsection{Comparison of HG Types}
To compare the assisting performance between HG types, we used the OAHG (UT), UPHG (UT+UA), and CombHG (UT+UA+SC) conditions, which showed the best assisting performance for each HG type in Section 5.1. Hereafter, the above three conditions will represent each HG type. A repeated measures ANOVA with a Greenhouse-Geisser correction was conducted to determine the effect of HG types (i.e., NHG, OAHG, UPHG, and CombHG) on each metric. As an exception, since mean disagreement cannot be calculated under the NHG condition, the analysis for mean disagreement was conducted only between the other three HG types. The ANOVA analysis revealed that the HG type has a significant effect on all metrics and statistical values of each metric are as follows: win rate (\(F_{3,57}=8.112\), \(p<.001\), \(\eta^2=.299\)), mean smash speed (\(F_{3,57}=5.740\), \(p=.002\), \(\eta^2=.232\)), defense rate (\(F_{3,57}=3.014\), \(p=.041\), \(\eta^2=.137\)), mean disagreement (\(F_{2,38}=13.586\), \(p<.001\), \(\eta^2=.417\)), helpfulness (\(F_{3,57}=8.614\), \(p<.001\), \(\eta^2=.312\)), naturalness (\(F_{3,57}=30.107\), \(p<.001\), \(\eta^2=.613\)), controllability (\(F_{3,57}=30.083\), \(p<.001\), \(\eta^2=.613\)), and comfort (\(F_{3,57}=7.108\), \(p=.001\), \(\eta^2=.272\)). Post hoc tests with a Bonferroni correction were conducted to figure out whether HG type pairs had a statistically significant difference, and the results are summarized in Table~\ref{tab:03}.

\begin{table*}[t]
  \caption{Comparisons showing statistically significant differences between HG types.}
  \label{tab:03}
  \begin{tabular}{lcclcc}
    \toprule
    \textbf{Metrics}& \textbf{Comparison}& \textbf{\(p\)}&
    \textbf{Metrics}& \textbf{Comparison}& \textbf{\(p\)}\\
    \midrule
    Win rate& OAHG \(>\) NHG& .014& Naturalness& NHG \(>\) OAHG& .000\\
     & UPHG \(>\) NHG& .002& & NHG \(>\) UPHG& .025\\
     & CombHG \(>\) NHG& .001& & UPHG \(>\) OAHG& .001\\
    Mean smash speed& CombHG \(>\) NHG& .023& & CombHG \(>\) OAHG& .000\\
     & CombHG \(>\) OAHG& .045& Controllability& NHG \(>\) OAHG& .000\\
    Defense rate& OAHG \(>\) NHG& .038& & UPHG \(>\) OAHG& .000\\
    Mean disagreement& OAHG \(>\) CombHG& .000& & CombHG \(>\) OAHG& .000\\
     & UPHG \(>\) CombHG& .002& Comfort& UPHG \(>\) OAHG& .003\\
    Helpfulness& UPHG \(>\) NHG& .009& & CombHG \(>\) OAHG& .001\\
     & CombHG \(>\) NHG& .001& & & \\
    \bottomrule
  \end{tabular}
\end{table*}

Based on the post hoc analysis, we can conclude the following: All three HG types, that is, OAHG, UPHG, and CombHG, led to significantly higher win rates of participants than NHG, which demonstrates the objective effectiveness of the HGs presented in this study. Specifically, CombHG induced a significantly higher mean smash speed than NHG, while OAHG induced a significantly higher defense rate than NHG. UPHG also induced a marginally higher smash speed than NHG, but it was not statistically significant (\(p=.073\)). Meanwhile, CombHG showed a lower mean disagreement than both UPHG and OAHG. In terms of the subjective metrics, UPHG and CombHG scored significantly higher in helpfulness than NHG, whereas OAHG did not. UPHG and CombHG received similar levels of subjective evaluation from participants, scoring significantly better than OAHG for the remaining three metrics, that is, naturalness, controllability, and comfort. NHG received the highest scores in naturalness and controllability, but this can be interpreted as an inevitable result of not exerting any artificial force on the users.

\section{Discussion}
We summarize the findings of this paper as answers to our research questions based on the analysis results and comments from the user interviews. In addition, we discuss the implications of our HG design regarding its generalization and utilization outside of the air hockey environment.

\subsection{Answers to Research Questions}

\textbf{RQ1:} \textit{Do the UT, UA, and SC methods that we propose and apply to the HG implementation contribute to an improvement in HG performance?}

The experimental results showed that the three methods proposed in this study contributed to the enhancement of the assisting performance in several metrics. First, the application of UT led to a reduction in mean disagreement for all HG types. This can be explained by the fact that the conflict between HG and a user because of an inaccurate model output, that is, with a high uncertainty, was effectively prevented by the uncertainty weight. The user interviews provided more details. Five participants (P6, P7, P11, P12, and P17) commented on OAHG (vanilla) that ``\textit{The interference of the guidance was excessive,}'' but only two participants (P8, P11) made such comments on OAHG (UT). A similar tendency was observed for UPHG. When comparing cases of non-applied and applied UT (i.e., UPHG (UA) vs. UPHG (UT+UA)), the number of participants commenting that ``\textit{The interference frequency of the guidance was appropriate}'' increased from zero to five (P7, P10, P16, P17, and P20).

On the other hand, the application of UA improved the users' evaluation of helpfulness and controllability. Moreover, when UA worked with UT on UPHG, there were enhancements in win rate, naturalness, and comfort when compared to UPHG (vanilla). This enhancement can be explained by UA providing the effect of adjusting HG to suit an individual user's playstyle, as we intended. In the interviews, three participants (P7, P8, and P13) reported on UPHG (UA) that, ``\textit{The guidance understood my offensive and defensive intentions,}'' whereas no one reported this on UPHG (vanilla).

The application of SC reduced the mean disagreement, which is similar to the effect of UT. This can be explained by the fact that the HG was fully delivered to the user only when OAHG and UPHG have a matched direction, thereby reducing the degree of HG interference. Note that there was no deterioration in other metrics, even though the HG interference was controlled this way. Rather, when SC worked with UT on CombHG (i.e., CombHG (UT+UA+SC)), the mean smash speed and all subjective evaluations were enhanced when compared to CombHG (UA). This suggests that users may prefer to reduce such unnecessary HG interference. A remark from P3 to CombHG (SC+UA), ``\textit{It felt positive that the guidance assisted me only when I needed it,}'' can supplement this.

Furthermore, we analyzed the distribution of \(W_{unc}\) and \(W_{sim}\) used in the UT and SC methods to determine the degree to which each method adjusted the guiding force to improve the HG performance. First, we calculated the mean and STD of the distribution of each weight within the same game. After averaging the mean and STD values over all games executed, the distribution (mean \(\pm\) STD) of \(W_{unc}\) for OAHG and UPHG were 0.600 \(\pm\) 0.390 and 0.751 \(\pm\) 0.306, respectively. Because a lower weight implies a stronger decrease in guiding force, the result indicated that the UT method more aggressively controlled the guiding force for OAHG than for UPHG. Meanwhile, the distribution of \(W_{sim}\) was measured as 0.515 \(\pm\) 0.343.

\textbf{RQ2:} \textit{Do OAHG, UPHG, and CombHG improve users' task performance when compared to NHG?}

In the user experiment, all three types of HG led to a significantly higher win rate for users than NHG, which clearly indicates that the user's haptic task performance was improved by the guiding force. The specific differences in assisting performance between the HG types are discussed in more detail in RQ3 and RQ4.

\textbf{RQ3:} \textit{What differences do OAHG and UPHG have in users' objective and subjective evaluations?}

There was no significant difference in participants' win rates between OAHG and UPHG. However, differences between the two HG types were revealed in other objective metrics. OAHG induced a significant increase in defense rate, whereas UPHG induced a marginal increase in mean smash speed when compared to NHG. This tendency was also observed in the user interviews. Six participants (P1, P2, P3, P5, P6, and P19) commented on OAHG that ``\textit{The guidance helped to defend the puck toward our goal,}'' which is the highest number among the other HG types. P6 additionally mentioned that ``\textit{It pinpointed the spot to be blocked for important defense,}'' which indicates that OAHG reinforced the user's insufficient defensive ability. On the other hand, the most frequent comment participants (P1, P3, P4, P7, P10, P13, P15, and P20) mentioned about UPHG was that ``\textit{The guidance assisted me in the direction I am moving,}'' which may induce an increase in the smash speed. This can be explained as the trained user prediction model could easily expect the user to continue to move in a direction when he/she starts moving, thereby providing an assisting guidance in that direction. Furthermore, P3 remarked that, ``\textit{It informed me where to stop smashing during the smashing process, so I was able to smash stable,}'' which indicates that UPHG assisted the user's smashing motion rather than simply accelerating in the moving direction.

In subjective metrics, user evaluation of OAHG and UPHG exhibited a clear difference. UPHG was evaluated to be significantly higher in most metrics, that is, in naturalness, controllability, and comfort, and marginally higher in helpfulness, when compared to OAHG. We looked for the cause of this one-sided subjective evaluation in the user interviews. Participants mentioned several negative comments on OAHG as follows: ``\textit{In an offensive situation, the guidance over-asserted its intention}'' (P5, P7, P9, P11, and P17); ``\textit{It followed the puck too hard}'' (P3, P10, P11, P13, and P15); ``\textit{It did not fit my intention sometimes}'' (P13, P14, P19, and P20). P3 remarked in detail, ``\textit{When the puck was on the opponent's side, I wanted to wait, but the guidance preferred to move from side to side along the puck.}'' These comments indicate that, even if the behavior suggested by OAHG is optimal, it could inconvenience the users when it does not match their intentions. On the other hand, it can be seen that UPHG received higher subjective evaluations in that it did not harm the intention of the participants. P7 remarked on UPHG, ``\textit{When I wanted to stay still, I could stay still, and when I tried to move toward the puck, I was assisted,}'' and five participants (P7, P10, P16, P17, and P20) also positively commented that ``\textit{The interference frequency of the guidance was appropriate.}'' Meanwhile, there was also a skeptical view on UPHG. P13 mentioned that ``\textit{I tried to stop in front of the puck, but as I was assisted in the direction of movement, it moved further and touched the puck,}'' indicating that an occasional inaccurate guidance could induce user's inconvenience.

\textbf{RQ4:} \textit{Can CombHG, which integrates OAHG and UPHG, complement each HG or provide better effects in users' objective and subjective evaluations?}

CombHG significantly lowered the mean disagreement than OAHG and UPHG, without reducing other objective and subjective metrics. This implies that CombHG succeeded in assisting users with less interference by effectively combining OAHG and UPHG. In detail, CombHG received a positive comment from six participants (P1, P3, P6, P13, P16, and P20), the highest number along with OAHG, that ``\textit{The guidance helped to defend the puck toward our goal.}'' Additionally, it also scored high subjective evaluations along with UPHG. Several user comments can summarize the advantages of CombHG. P3 remarked on CombHG that ``\textit{The guidance was actively helping me in the defensive situation and allowing me attack freely in the offensive situation, which felt ideal for me to play.}'' P19 also mentioned, ``\textit{I felt like the guidance was chasing the puck but giving me a choice to attack.}'' Through these interviews, we judged that CombHG satisfied participants by providing adequate guidance in situations where it is necessary (e.g., defense against the fast-approaching puck) and ensuring participants' autonomy in situations where various strategies are possible.

\subsection{Generalization}
The proposed OAHG and UPHG implementation methods (i.e., UT and UA) are applicable to other pHRI tasks if the optimal action and user behavior for the given task can be modeled as any neural network structure that receives the task state as input and outputs the action distribution (e.g., Figure~\ref{fig:03}). Because the UT method utilizes \(W_{unc}\) calculated using Equation (\ref{eq:05}) based on the output STD from the model, it is applicable to any neural network structure that outputs the STD of the action distribution. Because the UA method is based on the parameter update of the MAML with the user demonstration data (Equation (\ref{eq:01})), it is applicable to any neural network structure that can be trained by the MAML algorithm. Therefore, our framework is not limited to the video game, but it can be easily extended to other pHRI tasks because data-driven modeling of the optimal action or user behavior has been successfully demonstrated in various HG scenarios (e.g., robot-assisted surgery~\cite{power2015cooperative, zahedi2017gesture} and steering task~\cite{scobee2018haptic}). To model the optimal action, we can either apply reinforcement learning in simulated environments, which is not limited to the self-play-based learning, or use the movement of skillful experts as in~\cite{bluteau2008haptic, perez2016using, teranishi2017effects}. To model the user behavior, we can apply the MAML algorithm utilizing sufficient motion data from multiple users.

In addition, our CombHG implementation method (i.e., SC) is further general because it is not limited to the specific implementation methods of OAHG and UPHG. Any OAHG and UPHG implementations (e.g., the traditional OAHG method using a fixed reference path~\cite{forsyth2005predictive}) can be combined by the SC method because it requires only the similarity of guiding force vectors from OAHG and UPHG.

\section{Conclusion}
In this paper, we proposed deep learning-based novel implementation methods for OAHG and UPHG, applying a self-play-based reinforcement learning framework for OAHG and a meta-learning framework for UPHG to achieve their best performance. Further, we proposed CombHG that aimed to complement each HG type and provide better performance than OAHG and UPHG. In detail, the three proposed implementation methods (i.e., UT, UA, and SC) were applied to the given problem and demonstrated clear performance enhancement. Through the user study, we validated the assisting performance of each HG for users conducting a haptic task and investigated the difference in the user's subjective evaluation for each HG. The user study results indicated that UPHG and CombHG elicited significantly better subjective scores than OAHG. In addition, CombHG exhibited a further decrease in user disagreement compared to OAHG and UPHG, without reducing any objective and subjective scores. The comparison of each HG type based on our experimental analyses and user interviews can suggest the criteria for general HG design based on the aspects of HG that positively or negatively affect users. Considering that the generalization of the proposed HG implementation methods for other HG applications is straightforward, our findings are expected to contribute to the design of other HG-based pHRI applications beyond the video game environment considered in this study.

\begin{acks}
This work was supported by the Basic Science Research Program through the National Research Foundation of Korea (NRF) funded by the Ministry of Education (NRF-2018R1D1A1B07043580).
\end{acks}
\balance{}

\bibliographystyle{ACM-Reference-Format}
\bibliography{references.bib}

\end{document}